\documentclass[12pt]{article}
\usepackage{amsmath,amssymb,epsfig}
\makeatletter \@addtoreset{equation}{section} \makeatother
\renewcommand{\theequation}{\thesection.\arabic{equation}}

\newcommand{\ba}{\begin{array}}
\newcommand{\ea}{\end{array}}
\newcommand{\beq}{\begin{equation}}
\newcommand{\eeq}{\end{equation}}
\newcommand{\bea}{\begin{eqnarray}}
\newcommand{\eea}{\end{eqnarray}}

\def\bce{\begin{center}}
\def\ece{\end{center}}

\def\nonu{\nonumber}

\def\pa{\partial}
\def\al{\alpha}
\def\be{\beta}

\def\th{\theta}

\def\eps6{{\displaystyle \mathop{\epsilon}^{6}}{}}

\def\nab6{{\displaystyle \mathop{\nabla}^{6}}{}}

\def\ft#1#2{{\textstyle{\frac{\scriptstyle #1}{\scriptstyle #2}}}}
\def\fft#1#2{\frac{#1}{#2}}
\def\0{{\sst{(0)}}}
\def\1{{\sst{(1)}}}
\def\2{{\sst{(2)}}}
\def\3{{\sst{(3)}}}
\def\4{{\sst{(4)}}}
\def\5{{\sst{(5)}}}
\def\6{{\sst{(6)}}}
\def\7{{\sst{(7)}}}
\def\8{{\sst{(8)}}}

\def\td{\tilde}
\def\nnn{\nonumber}
\def\CP{{{\mathbb C}{\mathbb P}}}

\def\ba{\begin{array}}
\def\ea{\end{array}}
\def\beq{\begin{equation}}
\def\eeq{\end{equation}}
\def\be{\begin{equation}}
\def\ee{\end{equation}}

\def\eps{\epsilon}

\def\th{{\theta}}

\def\ba{\begin{array}}
\def\ea{\end{array}}
\def\beq{\begin{equation}}
\def\eeq{\end{equation}}
\def\be{\begin{equation}}
\def\ee{\end{equation}}

\def\eps{\epsilon}

\def\th{{\theta}}

\newcommand{\bean}{\begin{eqnarray*}}
\newcommand{\eean}{\end{eqnarray*}}

\begin{document}
\thispagestyle{empty} \addtocounter{page}{-1}
\begin{flushright}
{\tt hep-th/0505168}\\
\end{flushright}

\vspace*{1.3cm} \centerline{ \Large \bf Marginal Deformations with
$U(1)^3$ Global Symmetry } \vspace*{1.5cm} \centerline{{\bf
Changhyun Ahn$^{1,2}$
and {\bf Justin F. V\'azquez-Poritz}$^{1,3}$}} \vspace*{1.0cm}
\centerline{\it  $^{1}$ School of Natural Sciences, Institute for
Advanced Study, Einstein Drive, Princeton NJ 08540, USA}
\centerline{\it $^{2}$ Department of Physics, Kyungpook National
University, Taegu 702-701, Korea} \centerline{\it $^{3}$
Department of Physics, University of Cincinnati, Cincinnati OH
45221-001, USA } \vspace*{0.8cm} \centerline{\tt ahn@ias.edu,
\qquad jporitz@ias.edu} \vskip2cm

\centerline{\bf Abstract} \vspace*{0.5cm}

We generate new 11-dimensional supergravity solutions from
deformations based on $U(1)^3$ symmetries. The initial geometries
are of the form AdS$_4\times Y_7$, where $Y_7$ is a 7-dimensional
Sasaki-Einstein space. We consider a general family of
cohomogeneity one Sasaki-Einstein spaces, as well as the
recently-constructed cohomogeneity three $L^{p,q,r,s}$ spaces. For
certain cases, such as when the Sasaki-Einstein space is ${\bf S}^7$,
$Q^{1,1,1}$ or $M^{1,1,1}$, the deformed gravity solutions
correspond to a marginal deformation of a known dual gauge theory.

\baselineskip=18pt
\newpage
\renewcommand{\theequation}
{\arabic{section}\mbox{.}\arabic{equation}}

\section{Introduction}

A marginal deformation of ${\cal N}=4$ super Yang-Mills theory to
${\cal N}=1$ theory preserving $U(1) \times U(1)$ global symmetry,
along with a $U(1)_R$ symmetry, provides a new type IIB
supergravity background geometry \cite{LM} via the AdS/CFT
correspondence \cite{Malda1997,GKP,Witten1998}. The gravity dual
of the original undeformed theory  has an isometry group which
includes $U(1) \times U(1)$. The marginal deformation of the gauge
theory can be described by an $SL(2,R)$ transformation acting on a
2-torus ${\bf T}^2$ in the gravity solution. This particular
$SL(2,R)$ transformation produces a non-singular geometry provided
that the original geometry is non-singular. Gravity duals
corresponding to marginal deformations of other field theories
based on conifolds and toric manifolds are also given in
\cite{LM}.

A prescription for finding the marginal deformations of
eleven-dimensional gravity solutions with $U(1)^3$ global symmetry
was also provided in \cite{LM}, and was applied to the case of
AdS$_4\times {\bf S}^7$. The isometry group of ${\bf S}^7$ is
$SO(8)$. The $U(1)^3$ symmetries can be embedded in the $SU(4)$
subgroup of $SO(8)$, which implies that the deformed solution
preserves two supersymmetries in three dimensions. With the
appropriate coordinates, the angular directions corresponding to
the three $U(1)$'s can be identified, and the metric can be
written in a form that explicitly shows the three-torus ${\bf
T}^3$ symmetry. An additional angle is related to the
$SO(2)_R=U(1)_R$ $R$-symmetry, which is a symmetry of the usual
3-dimensional ${\cal N}=2$ superconformal field theories. Two of
the ${\bf T}^3$ angles are used to dimensionally reduce and
T-dualize the solution to type IIB theory.
Performing an $SL(2,R)$ transformation and then T-dualizing and
lifting back to eleven dimensions on the transformed directions
yields a new 11-dimensional solution. This deformed solution has a
warp factor, as well as an additional term in the 4-form field
strength, which depends on the deformation parameter. In the limit
of vanishing deformation parameter, the original AdS$_4\times {\bf
S}^7$ solution is regained.

This prescription can be readily applied to other 11-dimensional
solutions with geometries of the form AdS$_4\times Y_7$ provided
that, in addition to the R-symmetry group, the isometry group
contains $U(1)^3$. Although the operator that we are adding to the
Lagrangian of the corresponding three-dimensional dual gauge
theory is not known, we expect that this is a marginal deformation
because the deformed theory is conformal. We will consider cases
for which $Y_7$ is a 7-dimensional Sasaki-Einstein space. Then the
initial 3-dimensional dual gauge has two supersymmetries and the
deformation does not break any supersymmetry. A $d$-dimensional
Sasaki-Einstein space can be defined as the Einstein base of a
Calabi-Yau cone, and can always be written in canonical form as a
$U(1)$ bundle over an Einstein-K\"{a}hler metric. This $U(1)$
corresponds to the R-symmetry of the gauge theory. The requirement
that the global symmetry group of the gauge theory includes
$U(1)^3$ corresponds to the condition that the $U(1)^3$ lies
within the isometry group of the Einstein-K\"{a}hler base space.
The isometry groups of most of the Sasaki-Einstein spaces we
consider have $SU(2)$ or $SU(3)$ elements, which contain $U(1)$
and $U(1)^2$ subgroups, respectively. Thus, the resulting deformed
spaces have isometry groups in which the $SU(2)$ and $SU(3)$
factors are replaced by $U(1)$ and $U(1)^2$ accordingly.

Until recently, few explicit metrics were known for
Sasaki-Einstein spaces. A countably infinite number of
5-dimensional Sasaki-Einstein manifolds of topology ${\bf S}^2
\times {\bf S}^3$ has been constructed in
\cite{GMSW03-1,GMSW03-2}\footnote{The ${\cal N}=1$ superconformal
gauge theories living on the worldvolume of the D3-branes, dual to
5-dimensional anti-de Sitter space times 5-dimensional
Sasaki-Einstein manifolds, were constructed in \cite{BFHMS}. See
also the relevant works in
\cite{GMSW03-2,HSY,GMSW11,CLPP,MS,BBC1,HEK,LT,BHK,Pal,Gauntlett2005,BH,FHU,HKW,MSY,Frolov,BLMZ,MS1,BHOP,GP,BK,BBC,FHSU,DKR,GN}.}.
These $Y^{p,q}$ spaces are characterized by two coprime
positive integers $p$ and $q$. The marginal deformations of type
IIB solutions with geometries of the form AdS$_5\times Y^{p,q}$
have already been considered in \cite{LM}.

Higher-dimensional Sasaki-Einstein spaces were found in
\cite{GMSW03}. We will focus on the 7-dimensional spaces, which we
refer to as $X^{p,q}$. These spaces are cohomogeneity one and
include the previously-known homogeneous spaces ${\bf S}^7$,
$Q^{1,1,1}$ and $M^{1,1,1}$ as special cases\footnote{A
homogeneous space has an isometry group $G$ which is transitive.
In this case, the manifold can be expressed as a coset space
$G/\Gamma$, where $\Gamma$ is the isotropy group. On the other
hand, when generic orbits have $n$ dimensions less than the
dimensionality of the space, then the space is said to be
cohomogeneity $n$.}. The 6-dimensional Einstein-K\"{a}hler base
space of $X^{p,q}$ can be expressed as a 2-dimensional bundle over
a 4-dimensional Einstein-K\"{a}hler space $B_4$. We have a couple
of choices for the base space $B_4$, namely $\CP^2$ or
$\CP^1\times \CP^1$. For $B_4=\CP^2$, ${\bf S}^7$ and $M^{1,1,1}$
arise as particular cases while, for $B_4=\CP^1\times \CP^1$,
$Q^{1,1,1}$ arises as a special case. The $X^{p,q}$ family of
spaces can be further generalized for $B_4=\CP^1\times \CP^1$ by
rendering the characteristic radii of the two 2-spheres to be
different \cite{GMSW11,CLPP}. This is also a cohomogeneity one
family of Sasaki-Einstein spaces. We will refer to these more
general spaces as $Z^{p,q,r,s}$, which are characterized by four
positive coprime integers $p$, $q$, $r$ and $s$.

It has been found that the 5-dimensional $Y^{p,q}$ spaces can be
generalized to cohomogeneity two spaces $L^{p,q,r}$
\cite{Cvetic:2005ft}. This was found by noting that the connection
between Sasaki-Einstein spaces and special BPS scaling limit of
Euclideanized equal-angular-momenta Kerr-de Sitter black holes
\cite{HSY,CGS} 
could be generalized to the case in which there are two
different angular momenta. Likewise, the 7-dimensional $X^{p,q}$
with $B_4=\CP^2$ can be generalized to cohomogeneity three spaces
$L^{p,q,r,s}$, which is related to 7-dimensional Kerr-de Sitter
black holes with three independent angular momenta.

This paper is organized as follows. In section 2, we find the
marginal deformation of AdS$_4\times Q^{1,1,1}$. Next, we apply
this deformation procedure to 11-dimensional geometries which
include various 7-dimensional Sasaki-Einstein spaces. In section
3, we consider the infinite family of cohomogeneity one
Sasaki-Einstein spaces $Z^{p,q,r,s}$, which include the spaces
${\bf S}^7$ and $Q^{1,1,1}$ as particular cases. We provide a
brief overview of these spaces in subsection 3.1 and apply the
deformations to them in subsection 3.2. In section 4, we consider
the $M^{1,1,1}$ space (subsection 4.1) and its cohomogeneity three
generalization $L^{p,q,r,s}$ (subsection 4.2), which also includes
${\bf S}^7$ as a particular case. A number of possible further
directions are presented in section 5, including the application
of this deformation procedure to other types of geometries. For
the convenience of the readers, we have included a compendium of
Sasaki-Einstein spaces in Appendix A, which discusses how the
various families of these spaces are related. Flow diagrams
explicitly show how families of Sasaki-Einstein spaces are related
in various limits, for both the 5 and 7-dimensional cases. We
include some calculational details of the deformation of
AdS$_4\times L^{p,q,r,s}$ in Appendix B. Lastly, although the main
theme of our paper is marginal deformations of 11-dimensional
solutions, we have included the deformation of the type IIB
solution AdS$_5\times L^{p,q,r}$ in Appendix C, since the
recently-obtained 5-dimensional $L^{p,q,r}$ spaces have not yet
been discussed in this context.

\section{Deforming AdS$_4 \times Q^{1,1,1}$}

The simplest examples of 7-dimensional Sasaki-Einstein spaces
include the round ${\bf S}^7$, $Q^{1,1,1}$ and $M^{1,1,1}$. Since
the case of ${\bf S}^7$ has already been done in \cite{LM}, here
we begin by considering $Q^{1,1,1}$. By putting a large number of
$N$ coincident M2-branes at a conical singularity and taking the
near-horizon limit, we obtain the 11-dimensional solution
\cite{Fabbri1,Fabbri2,GNS,Ahn2002,Ahn1999} (See also
\cite{NP,PP1,PP2})
\bea
d s_{11}^2 = \frac{1}{4} d s_{{\rm AdS}_4}^2 + ds_{Q^{1,1,1}}^2,
\qquad F_{(4)} = \frac{3}{8} \left(\frac{3}{8} \right)^{1/6}
\omega_{AdS_4}\,,
\label{eleven}
\eea
where $\omega_{AdS_4}$ is the volume element of a unit AdS$_4$
spacetime. The four-form field strength has an extra $
\left(\frac{3}{8} \right)^{1/6}$ factor compared with the case of
the round 7-sphere, which arises from the ratio of the volumes of
$Q^{1,1,1}$ and ${\bf S}^7$. The metric for $Q^{1,1,1}$ can be
written as
\bea
ds_{Q^{1,1,1}}^2 &=& \frac{1}{64} \left( d \psi +
\sum_{i=1}^{3} c_{\th_i} d \phi_i \right)^2 + \frac{1}{32}
\sum_{i=1}^{3} \left(  d \theta_i^2 + s_{\th_i}^2 d \phi_i^2
\right)\,.
\label{sevenmetric}
\eea
Topologically, $Q^{1,1,1}$ is a $U(1)$ bundle over ${\bf S}_1^2
\times {\bf S}_2^2 \times {\bf S}_3^2$. The base is parameterized
by the spherical coordinates $(\th_i, \phi_i)$ where $i=1,2,3$
denotes the $i^{th}$ 2-sphere\footnote{We use the simple notation
$c_{\th_i} \equiv \cos \th_i$ and $s_{\th_i} \equiv \sin \th_i$.}
and the angle $\psi$ parameterizes the $U(1)$ Hopf fiber. The
$SU(2)_1 \times SU(2)_2 \times SU(2)_3 \times U(1)$ isometry group
of $Q^{1,1,1}$ corresponds to the $SU(2)_1 \times SU(2)_2 \times
SU(2)_3$ global symmetry and $U(1)$ $R$-symmetry of the dual
conformal field theory of \cite{Fabbri2}.

We will now apply the deformation procedure of \cite{LM} to the
above 11-dimensional supergravity solutions\footnote{We follow the
notations from \cite{LM}.}. With regards to the required $U(1)^3$
symmetry, one $U(1)$ can be taken from each $SU(2)$ of the
isometry group. Therefore, this deformation preserves the $U(1)$
$R$-symmetry. Since the original dual gauge theory is ${\cal N}=2$
supersymmetric, the deformed gauge theory will also be ${\cal
N}=2$ supersymmetric. In other words, the three $U(1)$'s commute
with the ${\cal N}=2$ supercharge.

The 11-dimensional metric can be decomposed into a 3-dimensional
piece, for which the ${\bf T}^3$ symmetry is explicit, and a
remaining 8-dimensional piece, as given in (A.1) of \cite{LM}. In
rewriting the above 11-dimensional metric in the form of (A.1), we
take the ``$11^{th}$'' direction as the $\phi_3$ direction. By
reading off the $d\phi_3^2$ terms from (\ref{sevenmetric}) one
gets
\be \Delta^{1/3} e^{4\phi/3} = \fft{1}{64}(c_{\theta_3}^2
+2s_{\theta_3}^2)\,. \nonu \ee
The warp-factor $\Delta$ is the determinant of the $3\times 3$
matrix represented by the metric for $D\phi_i$ and is given by
\bea
\Delta(\th_1,\th_2,\th_3) &= & \frac{c_{\th_3}^2 s_{\th_1}^2
s_{\th_2}^2 - \frac{1}{2}( -2 + c_{2\th_1} + c_{2\th_2})
s_{\th_3}^2}{65536}\,.
\nonu
\eea
>From these two results, one can read off $e^{4\phi/3}$. From the
expression of (\ref{sevenmetric}), one can decompose the
3-dimensional metric into a 2-dimensional metric for the subspace
$(\phi_1,\phi_2)$ and another term which takes the form
\bea
\Delta^{1/3} e^{4\phi/3} \left( D \phi_3 + N_1 D \phi_1 +N_2
D \phi_2 \right)^2\,.
\label{three}
\eea
Now one can read off the term $d \phi_3 d \phi_1$ and the term $d
\phi_3 d \phi_2$ which will determine $N_1$ and $N_2$ of
\cite{LM}:
\bea
N_1 =-\frac{2 c_{\th_1} c_{\th_3}}{(-3+ c_{2\th_3})}, \qquad
N_2 =- \frac{2c_{\th_2} c_{\th_3}}{(-3+c_{2\th_3})}\,.
\nonu
\eea

Clearly, the 2-dimensional metric from the 11-dimensional point of
view can be written as
\bea
&& \Delta^{1/3} e^{-2\phi/3} h_{mn} D \phi^m D \phi^n  =
\frac{(-2+c_{2\theta_1}+c_{2\theta_3})}{32(-3+c_{2\theta_3})}
\left( D\phi_1 + \frac{2 c_{\th_1} c_{\th_2}
s_{\th_3}^2}{-2+c_{2\th_1}+c_{2\th_3}}
D\phi_2 \right)^2
\label{two} \\
&&- \frac{c_{\th_3}^2(-2+c_{2\th_1}+c_{2\th_3})s_{\th_2}^2 +
(c_{\th_2}^2(-3+c_{2\th_3}) s_{\th_1}^2 + 2(-2+c_{2\th_1}
+c_{2\th_3}) s_{\th_2}^2)s_{\th_3}^2}
{16(-3+c_{2\th_3})(-2+c_{2\th_1}+c_{2\th_3})} D\phi_2^2\,,
\nnn
\eea
where we introduce the following notation $D \phi_i = d \phi_i +
{\cal A}^i$ and the fields
\bea {\cal A}^1 &=&\frac{8c_{\theta_1} s_{\theta_2}^2
s_{\theta_3}^2}{H}  d \psi\,,
\nonu \\
{\cal A}^2 &=& \frac{1}{\left[
-\frac{1}{2}(-2+c_{c_{2\th_1}}+c_{2\th_2}
  ) \frac{1}{s_{\th_1}^2 c_{\th_2}}
+\frac{s_{\th_2}^2 c_{\th_3}^2}{c_{\th_2} s_{\th_3}^2} \right] } d
\psi\,, \nonu \\
{\cal A}^3 &=& \frac{8c_{\theta_3} s_{\theta_1}^2
s_{\theta_2}^2}{H}  d \psi\,. \nonu \eea
where
\be H=5-3c_{2\theta_3}+c_{2\theta_1}(-3+c_{2\theta_2}
+c_{2\theta_3})+ c_{2\theta_2}(-3+2c_{\theta_1}^2
c_{2\theta_3})\,. \ee
One can read off the metric $h_{mn}$, whose determinant is equal
to 1, from (\ref{two}). The fields $\phi^i$, ${\cal A}^i$ and
three-form gauge fields transform appropriately under the general
$SL(3,R)$ transformation of coordinates $\phi^i$.

The remaining 8-dimensional metric is given by
\bea
\Delta^{-1/6} g_{\mu \nu} dx^{\mu} dx^{\nu} &= & \frac{1}{4}
d s_{AdS_4}^2 + \frac{1}{32} \left( d \theta_1^2 + d \theta_2^2 +
d \theta_3^2 \right)
\label{eight}
\\
&+& \frac{s_{\th_1}^2 s_{\th_2}^2 s_{\th_3}^2}{ 4H} d \psi^2\,.
\nnn \eea
The 11-dimensional metric (\ref{eleven}) can now be written in a
form which makes the ${\bf T}^3$ more explicit by adding together
(\ref{three}), (\ref{two}), and (\ref{eight}).

We now dimensionally reduce this solution along the $\phi_3$
direction and T-dualize along the $\phi^1$ direction. The
resulting type IIB solution can be written in the form of (A.5)
and (A.6) in \cite{LM}:
\bea
ds_{IIB}^2 & = & \frac{1}{h_{11}}
\frac{d\phi_1^2}{\sqrt{\Delta}} + \frac{\sqrt{\Delta}}{h_{11}}
D\phi_2^2 + e^{2\phi/3} g_{\mu \nu} dx^{\mu} dx^{\nu}\,,
\nonu \\
B &= &  \left( \frac{2 c_{\th_1} c_{\th_2}
s_{\th_3}^2}{-2+c_{2\th_1}+c_{2\th_3}} \right) d\phi_1 \wedge D
\phi_2 + d \phi_1 \wedge {\cal A}^1\,,
\nonu \\
e^{2\Phi} & =& \frac{e^{2\phi}}{h_{11}}\,, \nonu \\
C^{(0)} & = & -\frac{2 c_{\th_1} c_{\th_3}}{(-3+ c_{2\th_3})}\,,
\nonu \\
C^{(2)} & =& \left( \frac{2c_{\th_2} c_{\th_3}s_{\th_1}^2} {-2+
c_{2\th_1} +c_{2\th_3}} \right) d \phi_1 \wedge D \phi_2 -d \phi_1
\wedge {\cal A}^3\,,
\nonu \\
C^{(4)} & = & -\left( \frac{3}{8} \right)^{7/6} \left( \omega_3
\wedge d \phi_1 + \ast_8\omega_3 \wedge d \phi_2 \right)\,,
\label{IIBsol} 
\eea
where $\ast_8$ is the Hodge dual relative to the 8-dimensional
metric $g_{\mu\nu}$ and $d\omega_3=\omega_{AdS_4}$.

We will now take a particular $SL(2,R)$ transformation which will
yield a regular deformed 11-dimensional solution, provided that
the initial solution is regular \cite{LM}. The $SL(2,R)$
transformation is parameterized by ${\hat\gamma}\equiv
R^3\,\gamma$ in the $\phi_1-\phi_2$ plane. The modified
warp-factor and three-form \cite{LM} are given by
\bea
\Delta' = \frac{\Delta}{(1 + \hat{\gamma}^2 \Delta)^2} \equiv
G^2 \Delta, \qquad C_{123}' = -\frac{\hat{\gamma}\Delta}{1+
\hat{\gamma}^2 \Delta}\,,
\nonu
\eea
with other quantities unchanged.

T-dualizing and lifting along the transformed directions results
in the modified 11 dimensional solution
\bea
ds_{11}^2 & = & G^{2/3} \left[
\frac{(-2+c_{2\theta_1}+c_{2\theta_3})}{32(-3+c_{2\theta_3})}
\left( D\phi_1 + \frac{2 c_{\th_1} c_{\th_2}
s_{\th_3}^2}{-2+c_{2\th_1}+c_{2\th_3}}
D\phi_2 \right)^2 \right. \nonu \\
&-&
\frac{c_{\th_3}^2(-2+c_{2\th_1}+c_{2\th_3})s_{\th_2}^2 +
(c_{\th_2}^2(-3+c_{2\th_3}) s_{\th_1}^2 +
2(-2+c_{2\th_1} +c_{2\th_3}) s_{\th_2}^2)s_{\th_3}^2}
{16(-3+c_{2\th_3})(-2+c_{2\th_1}+c_{2\th_3})} D\phi_2^2  \nonu \\
&+& \left. \frac{(3-c_{2\th_3})}{128} \left( D \phi_3
-\frac{2 c_{\th_1} c_{\th_3}}{(-3+ c_{2\th_3})}
D \phi_1
- \frac{2c_{\th_2} c_{\th_3}}{(-3+c_{2\th_3})}
D \phi_2 \right)^2 \right] \nonu \\
&+& G^{-1/3} \left[ \frac{1}{4} d s_{{\rm AdS}_4}^2 + \frac{1}{32}
\left( d \theta_1^2 + d \theta_2^2 +
d \theta_3^2 \right) \right. \nonu \\
&+& \left. \frac{s_{\th_1}^2 s_{\th_2}^2 s_{\th_3}^2}{ 4H} d
\psi^2
\right]\,,\nnn\\
F_{(4)} & = & \left( \fft{3}{8} \right)^{7/6} \left(
\omega_{AdS_4} + \hat{\gamma} \sqrt{\fft{\Delta}{512H}}
s_{\th_1}s_{\th_2}s_{\th_3}\, d \th_1 d \th_2 d \th_3 d \psi
\right) - \hat{\gamma}\, d \left( \Delta G\, D \phi_1 D \phi_2 D
\phi_3 \right)\,. \label{11q111} \eea
While the initial geometry was the direct product AdS$_4\times
Q^{1,1,1}$, the above deformed geometry is a warped product of
these spaces. The warp factor $G$ depends on three of the internal
directions of $Q^{1,1,1}$. Since $G^{-1}\ge 1$, the deformed
geometry is guaranteed to be regular since the initial geometry
was regular.

Later examples of 11-dimensional deformed solutions given in this
paper will have a four-form field strength that has similar form
to the case above. The main difference will be in the $d \th_1 d
\th_2 d \th_3 d \psi$ term, which arises from
$\ast_8\,\omega_{AdS_4}$, where the Hodge dual is with respect to
a particular 8-dimensional space $g_{\mu\nu}$.

So far, we have described a particular 7-dimensional space
$Q^{1,1,1}$. This is a special case of the Einstein spaces
$Q^{p,q,r}$, where $p$, $q$ and $r$ represent the winding numbers
of the $U(1)$ bundle over the three 2-spheres. The $Q^{p,q,r}$
space has an isometry $SU(2)^3 \times U(1)$, except $Q^{0,0,1}$.
One can choose $U(1)^3$ as a subgroup of $SU(2)^3$. For the case
of $Q^{0,0,1}$, the isometry group is given by $SU(2)^4$.
Therefore, it is also possible to consider three $U(1)$'s. It is
straightforward to apply the above deformation procedure to all of
these spaces. However, the $Q^{1,1,1}$ space which we considered
above is the only case which is supersymmetric, with holonomy
group $SU(3)$.

\section{Cohomogeneity one generalization of $Q^{1,1,1}$}

\subsection{The $Z^{p,q,r,s}$ spaces}

We will now consider a class of cohomogeneity one Sasaki-Einstein
spaces $Z^{p,q,r,s}$, which contain $Q^{1,1,1}$ and ${\bf S}^7$ as
particular cases. A countably infinite number of 7-dimensional
Sasaki-Einstein spaces were found in \cite{GMSW03}. The
corresponding metrics can be expressed as a circle bundle over a
6-dimensional Einstein-K\"{a}hler base space which, in turn, is a
2-dimensional bundle over a 4-dimensional Einstein-K\"{a}hler base
space. For the case in which the 4-dimensional Einstein-K\"{a}hler
space is a direct product of two equal-radius 2-spheres, this
construction can be generalized such that the spheres have
different characteristic radii $\ell_1$ and $\ell_2$
\cite{GMSW11,CLPP}. The corresponding metric of these
Sasaki-Einstein spaces $Y_7$ expressed in canonical form is
\bea
ds_{Y_7}^2 & = &
\left[ d\psi^{\prime} +2A_{(1)}\right]^2+ds_{EK_6}^2\,.
\label{newsasakiform7}
\eea
where the 6-dimensional Einstein-K\"{a}hler metric is
\bea
ds_{EK_6}^2 &=& \frac{1}{8} (\ell_1-n_1\,y)(d \th_1^2
 + s_{\th_1}^2 d \phi_1^2) +
\fft18 (\ell_2-n_2\,y)(d \th_2^2 +s_{\th_2}^2 d
 \phi_2^2)\nnn\\ &+& \frac{1}{F(y)} dy^2 + \frac{F(y)}{64} (d\beta -
n_1\,c_{\th_1} d\phi_1 -n_2\, c_{\th_2} d \phi_2)^2\,
\label{EK6}
\eea
and
\be
A_{(1)}=\fft18 [\ell_1  c_{\th_1} d\phi_1 + \ell_2 c_{\th_2} d
\phi_2+ y (d \beta -n_1\,
 c_{\th_1} d\phi_1 -n_2\, c_{\th_2} d \phi_2)]\,.
\nonu
\ee
The K\"{a}hler form for $ds_{EK_6}^2$ is $J=dA_{(1)}$. The
function $F(y)$ is given by
\be F(y) = \fft{\fft43 a+\fft{16\ell_1
\ell_2}{n_1}(\ell_1-1)\,y-8\delta\, y^2+\fft{16}{3}
(n_1\ell_2-n_2+2n_2\ell_1)\,y^3 -4n_1 n_2\, y^4}{(\ell_1-n_1\,
y)(\ell_2-n_2\, y)} \,, \label{F} \ee
where $\delta\equiv \ell_2 (2\ell_1-1) +\ell_1 (\ell_1-1)n_2/n_1$.
It is obvious that the above metric (\ref{newsasakiform7}) has an
isometry $SU(2) \times SU(2) \times U(1) \times U(1)$. One can
obtain the above metric from the 7-dimensional Sasaki-Einstein
metric in (4) of \cite{CLPP} by taking
\bea
(\ell_1,\ell_2,p,q)\rightarrow
\fft18 (\ell_1,\ell_2,n_1,n_2),
\qquad r\rightarrow -y,
\qquad
\tau^{\prime}\rightarrow \fft18 \beta.
\nonu
\eea
For vanishing $a$, the
present $F(y)$ coincides with $c(r)^2$ in \cite{CLPP}. The value
for the integration constant $a$ in (\ref{newsasakiform7})
reflects a choice in the origin of $y$. In \cite{CLPP}, $a$ was
set to zero. Also, the metric has a rescaling symmetry under which
\bea
n_i\rightarrow \lambda\,n_i,
\qquad
a\rightarrow a/\lambda^2, \qquad
y\rightarrow y/\lambda, \qquad
\beta\rightarrow \lambda\,\beta.
\label{rescaling}
\eea
This implies that only the ratio $n_1/n_2$ of the parameters $n_1$
and $n_2$ is nontrivial. It has been found that the metric
(\ref{EK6}) is an Einstein-K\"{a}hler solution provided that the
algebraic constraint
\be
\fft{n_1}{n_2}=\fft{\ell_1-1}{\ell_2-1}\,
\label{constraint}
\ee
is satisfied. The constraint (\ref{constraint}) was not taken into
account in the local expressions given in \cite{GMSW11}. In
particular, if we neglect this constraint by taking $\ell_i=1$ and
keeping $n_1$ and $n_2$ as independent paramaters, then our $F(y)$
given in (\ref{F}) reduces to the expression given in
\cite{GMSW11}.

We have purposely kept repetitive parameters in order to more
clearly see limiting cases. In the limit $\ell_1=\ell_2=1$ and
$n_1=n_2\equiv c$, the space $Y_7$ reduces to the
seven-dimensional space found in \cite{GMSW03}. In this case, $a$
is now the nontrivial parameter. The limit
enlarging one of the two $U(1)$'s into $SU(2)$ symmetry
\bea \ell_i\rightarrow 1\,,\qquad n_i \rightarrow 0\,, \qquad y
\rightarrow c_{\th_3}\,, \qquad a \rightarrow 6\,, \qquad \beta
\rightarrow \phi_3\,,
\label{Qlimit}
\nonu
\eea
gives rise to the metric of $Q^{1,1,1}$, the case discussed in
previous section. On the other hand, in the limit
\bea \ell_i\rightarrow 1\,,\qquad n_1 \rightarrow 1\,,\qquad
n_2\rightarrow -1\,,\qquad a\rightarrow 3\,, \label{S7limit} \eea
the metric (\ref{EK6}) reduces to the Fubini-Study metric on
$\CP^3$, which is given by
\bea
ds_{\CP^3} &=& d\mu^2+\fft14 c_{\mu}^2\,(d\theta_1^2
+s_{\theta_1}^2d\phi_1^2) +\fft14 s_{\mu}^2\, (d\theta_2^2
+s_{\theta_2}^2d\phi_2^2)\nnn\\ &+& \fft14
s_{\mu}^2\,c_{\mu}^2\,
(d\beta -c_{\theta_1}d\phi_1+c_{\theta_2}d\phi_2)^2\,,
\label{CP3}
\eea
where we have used the coordinate transformation
$y=-c_{2\mu}$.
In this case, the Sasaki-Einstein metric (\ref{newsasakiform7})
becomes that of a round ${\bf S}^7$, where
\be A_{(1)}=\fft18 (s_{\mu}^2-c_{\mu}^2)\,d\beta+\fft14
c_{\mu}^2\, c_{\theta_1}d\phi_1+\fft14 s_{\mu}^2\,
c_{\theta_2}d\phi_2\,,
\label{Kahler}
\ee
and $J=dA_{(1)}$ is the K\"{a}hler form for $\CP^3$.

We take $y$ to lie in the range $y_1\le y\le y_2$. Different
coordinates have been used in order to find the regularity
conditions for this metric \cite{CLPP}. In particular, regularity
requires that
\be
\fft{n_2-n_1}{n_2}< \ell_1 < 1\,,\qquad 0<\ell_2 < 1\,,
\nonu
\ee
where we have chosen $0\le n_1\le n_2$. It has been found that the
regular metrics are specified by two rational numbers, or by four
integers $p$, $q$, $r$ and $s$. Therefore, these spaces can be
denoted by $Z^{p,q,r,s}$, in analogy with the 5-dimensional spaces
$Y^{p,q}$ \cite{MS}.

Consider the coordinate transformation
\bea \psi^{\prime}=\fft14 \psi, \qquad \beta=-\alpha+\psi\,.
\nonu
\eea
We will now suppose that $n_i$ is nonzero so that any such factors
can be absorbed by the coordinates. One can then express the
metric (\ref{newsasakiform7}) as
\bea ds_{Y_7}^2 & = & \fft18 (\ell_1-n_1\, y)(d \th_1^2 +
s_{\th_1}^2 d \phi_1^2)+ \fft18 (\ell_2-n_2\, y) (d \th_2^2
+s_{\th_2}^2 d \phi_2^2) + \fft{1}{F(y)} dy^2
\nnn\\
&+& \fft{F(y)}{16(F(y) + 4y^2)} (d \psi + \ell_1 c_{\th_1} d\phi_1
+ \ell_2 c_{\th_2} d \phi_2)^2
\nnn\\
&+& \frac{F(y) + 4y^2}{64} \Big[ d \al + \Big( 1-\fft{4
y}{F(y)+4y^2}\Big) \, d \psi \nnn\\ &+& \Big( n_1-\fft{4\ell_1\,
y}{F(y)+4y^2}\Big) \, c_{\th_1} d\phi_1 +\Big( n_2-\fft{4\ell_2\,
y}{F(y)+4y^2}\Big) \, c_{\th_2} d \phi_2\Big] ^2\,.
\nonu
\eea
In the case of vanishing $n_i$, it is more appropriate to use the
$Q^{1,1,1}$ metric given by (\ref{11q111}).

\subsection{Deforming AdS$_4\times Z^{p,q,r,s}$}

The 7-dimensional Sasaki-Einstein space $Z^{p,q,r,s}$ can be used
to construct a solution of 11-dimensional supergravity. This
solution has the geometry AdS$_4\times Z^{p,q,r,s}$ and is given
by
\be d s_{11}^2 = \frac{1}{4} d s_{AdS_4}^2 + ds_{Y_7}^2, \qquad
F_{(4)} =\frac{3}{8} \left(\frac{\mbox{vol} \; Z^{p,q,r,s} }
{\mbox{vol}\; {\bf S}^7} \right)^{1/6} \omega_{AdS_4} \,. \nonu
\ee
The $U(1)^3$ symmetry required for the deformation can be taken
from the $SU(2)\times SU(2)\times U(1)$ portion of the isometry
group of $Z^{p,q,r,s}$, which still leaves a $U(1)$ for the
R-symmetry. This means that the ${\cal N}=2$ supersymmetry of the
dual gauge theory will be preserved by the deformation. The metric
$ds_{Y_7}^2$ can be written in the following form which makes the
$U(1)^3$ symmetry explicit:
\bea ds_{11}^2 &=& \fft14 d s_{{\rm AdS}_4}^2+\fft18
(\ell_1-n_1y)\,d\theta_1^2 +\fft18 (\ell_2-n_2 y)\,d\theta_2^2
+\fft{1}{F(y)}dy^2 \label{Y7T3}
\\
&+& \fft{F(y)+4y^2}{64} [D\alpha +\Big( n_1-\fft{4\ell_1\,
y}{F(y)+4y^2}\Big) c_{\theta_1}\,D\phi_1 +\Big( n_2-\fft{4\ell_2\,
y}{F(y)+4y^2}\Big) c_{\theta_2}\,D\phi_2]^2 \nnn\\ &+&
\fft{g}{8f}\,D\phi_1^2 + f\,[D\phi_2+Kf^{-1}\ell_1\ell_2
c_{\theta_1} c_{\theta_2}\,D\phi_1]^2 + \fft{K}{8g}\,
(\ell_1-n_1y)(\ell_2-n_2 y)s_{\theta_1}^2
s_{\theta_2}^2\,d\psi^2\,, \nnn \eea
where the coefficient functions are
\bea K &=& \fft{F(y)}{16(F(y)+4y^2)}\,,\qquad f=\fft18 (\ell_2-n_2
y) s_{\theta_2}^2+K\ell_2^2 c_{\theta_2}^2\,,
\nnn\\
g &=& f(\ell_1-n_1 y)s_{\theta_1}^2 +K\ell_1^2 (\ell_2-n_2 y)
s_{\theta_2}^2 c_{\theta_1}^2\, \label{kfg} \eea
and the connections corresponding to $\phi^1, \phi^2$, and
$\alpha$ are given by \bea {\cal A}^1 &=& \fft{K}{g}\,\ell_1
(\ell_2-n_2 y) c_{\theta_1} s_{\theta_2}^2\, d\psi\,, \qquad
{\cal A}^2 = \fft{K}{g}\,\ell_2 (\ell_1-n_1 y) c_{\theta_2}
s_{\theta_1}^2\, d\psi\,,
\nonu \\
{\cal A}^3 &=& \Big( 1-\fft{4y}{F(y)+4y^2}\Big) \,d\psi-\Big(
n_1-\fft{4\ell_1\, y}{F(y)+4y^2}\Big) c_{\theta_1}\,{\cal A}^1
-\Big( n_2-\fft{4\ell_2\, y}{F(y)+4y^2}\Big) c_{\theta_2}\,{\cal
A}^2. \nnn \eea
We can read off the various ingredients of the metric (\ref{Y7T3})
which we use to reduce and T-dualize the solution to type IIB
theory:
\bea \Delta^{1/3} e^{4\phi/3} &=& \fft{F(y)+4y^2}{64}\,,\qquad
\Delta=\fft{F}{8192K}(g+8K^2\ell_1^2 \ell_2^2 c_{\theta_1}^2
c_{\theta_2}^2)\,,\nnn\\
N_1 &=& \Big( n_1-\fft{4\ell_1\, y}{F(y)+4y^2}\Big)
c_{\theta_1}\,,\qquad N_2=\Big( n_2-\fft{4\ell_2\,
y}{F(y)+4y^2}\Big) c_{\theta_2}\,. \label{resultq111} \eea

After applying an $SL(2,R)$ transformation, we T-dualize and lift
back up to eleven dimensions along the transformed coordinates.
The resulting 11-dimensional deformed solution has the metric
\bea
&& ds_{11}^2 = G^{-1/3}\,\Big[ \fft14 d s_{{\rm
AdS}_4}^2+\fft18 (\ell_1-n_1y)\,d\theta_1^2 +\fft18 (\ell_2-n_2
y)\,d\theta_2^2 +\fft{1}{F(y)}dy^2
\nnn\\
&& + \fft{K}{8g}\,
(\ell_1-n_1y)(\ell_2-n_2 y)s_{\theta_1}^2 s_{\theta_2}^2\,d\psi^2
\Big] \nnn\\
&& + G^{2/3}\,\Big[ \fft{F(y)+4y^2}{64} [D\alpha +\Big(
n_1-\fft{4\ell_1\, y}{F(y)+4y^2}\Big) c_{\theta_1}\,D\phi_1 +
\Big( n_2-\fft{4\ell_2\, y}{F(y)+4y^2}\Big)
c_{\theta_2}\,D\phi_2]^2
\nnn\\
&& + \fft{g}{8f}\,D\phi_1^2 +
f\,[D\phi_2+Kf^{-1}\ell_1\ell_2 c_{\theta_1}
c_{\theta_2}\,D\phi_1]^2 \Big] \,,
\label{finalresult}
\eea
where the coefficient functions $K,f$, and $g$ are given by
(\ref{kfg}). While the initial geometry was the direct product
AdS$_4\times Z^{p,q,r,s}$, the above deformed geometry is a warped
product of these spaces with the factor given by
\be G^{-1}=1+{\hat\gamma}^2\,\Delta\, \nonu \ee
and $\Delta$ is given by (\ref{resultq111}). The warp factor
$G$ depends on three
of the internal directions of $Z^{p,q,r,s}$. Since $G^{-1}\ge 1$, the
geometry (\ref{finalresult}) is regular since the initial geometry
was regular.

Note that the deformation of AdS$_4\times Q^{1,1,1}$ discussed in
the previous section cannot be obtained as a special case of
(\ref{finalresult}). This is because the $Q^{1,1,1}$ metric
(\ref{11q111}) was expressed in different coordinates, which
means that we performed the dimensional reductions along different
directions. On the other hand, the deformation of AdS$_4\times
{\bf S}^7$ arises from (\ref{finalresult}) in the limit
(\ref{S7limit}).

\section{$M^{1,1,1}$ and its cohomogeneity three generalization}

\subsection{Deforming AdS$_4\times M^{1,1,1}$}

We will now consider the deformation of an 11-dimensional solution
with the geometry \cite{PP2,NP,Pope84,CDF,DF,FFGT,Ahn0502}
\be ds_{11}^2 = \frac{1}{4} ds_{{\rm AdS}_4}^2+ds_{M^{1,1,1}}^2\,,
\qquad F_{(4)} = \frac{3}{8} \left(\frac{27}{128} \right)^{1/6}
\omega_{AdS_4}\,, \nonu \ee
where the homogeneous 7-dimensional Sasaki-Einstein space
$M^{1,1,1}$ has the metric
\bea ds_{M^{1,1,1}}^2 &=& \fft{1}{256} \left[d\tau+3\sin^2\mu\,
(d\psi+\cos\theta\,d\phi)+2\cos {\td\theta}\,d{\td\phi}\right]^2
+\fft{1}{32} (d{\td\theta}^2+\sin^2{\td\theta}\,d{\td\phi}^2)
\nnn\\
&+& \fft{3}{16} \left[d\mu^2+\fft14
\sin^2\mu\,\left(d\theta^2+\sin^2\theta\, d\phi^2+\cos^2\mu\,(d
\psi+\cos\theta\,d\phi)^2 \right)\right]\,. \nonu \eea
The corresponding ${\cal N}=2$ dual gauge theory in
three-dimensions is known \cite{Fabbri2}. Topologically,
$M^{1,1,1}$ space is a nontrivial $U(1)$ bundle parameterized by
$\tau$ over $\CP^2 \times {\bf S}^2$. The $SU(3) \times SU(2)
\times U(1)$ isometry group of $M^{1,1,1}$ corresponds to $SU(3)
\times SU(2)$ global symmetry and $U(1)_R$ $R$-symmetry  of the
dual conformal field theory \cite{Fabbri2}.

By taking the two $U(1)$'s parameterized by $\psi$ and $\phi$ from
$SU(3)$ group and one $U(1)$ parameterized by $\td\phi$ from
$SU(2)$ group, we have a $U(1)^3$ for the deformation while
leaving the $U(1)$ R-symmetry of the dual gauge theory untouched.
Thus, the ${\cal N}=2$ supersymmetry will be preserved. We can
rewrite the metric in a form which makes the $U(1)^3$ symmetry
explicit:
\bea ds_{11}^2 &=& \frac{1}{4} ds_{{\rm AdS}_4}^2
+\fft{3}{16}\left(d\mu^2+ \fft14
s_{\mu}^2\,d\theta^2\right)+\fft{1}{32}d{\td\theta}^2
+\fft{1}{64}g\,c_{\mu}^2\,s_{\td\theta}^2\,d\tau^2\
\nnn\\
&+& \fft{3}{256f} s_{\mu}^2 \left(D\psi+c_{\theta}\, D\phi+
2fc_{\td\theta} D{\td\phi}\right)^2
+\fft{3}{64}s_{\mu}^2\,s_{\theta}^2\,D\phi^2
+ \fft{f}{32g}\,D{\td\phi}^2 \nonu \eea
where
\be f^{-1}\equiv 4-s_{\mu}^2\,,\qquad g^{-1}\equiv 2c_{\mu}^2\,
c_{\td\theta}^2+(4-s_{\mu}^2)s_{\td\theta}^2\, \nonu \ee
and the gauge connections corresponding to $\phi, \td\phi$ and
$\psi$ are given by \be {\cal A}^1=0\,,\qquad {\cal
A}^2=g\,c_{\mu}^2\,c_{\td\theta}\,d\tau\,,\qquad {\cal
A}^3=f(1-2g\,c_{\mu}^2\,c_{\td\theta})^2\,d\tau\,. \nonu \ee
One can then read off the various ingredients which are used to
reduce and T-dualize the solution to type IIB theory:
\bea \Delta^{1/3}e^{4\phi/3} &=& \fft{3}{256f}s_{\mu}^2\,,\qquad
\Delta=\fft{9s_{\mu}^4 s_{\theta}^2}{2^{19}g}\,,\nnn\\
N_1 &=& c_{\theta}\,,\qquad N_2 = 2f\,c_{\td\theta}\,. \nonu \eea
After applying an $SL(2,R)$ transformation, we lift back up to
eleven dimensions and find the deformed solution
\bea ds_{11}^2 &=& G^{-1/3} \Big[ \frac{1}{4} ds_{{\rm AdS}_4}^2
+\fft{3}{16}\left(d\mu^2+ \fft14
s_{\mu}^2\,d\theta^2\right)+\fft{1}{32}d{\td\theta}^2
+\fft{1}{64}g\,c_{\mu}^2\,s_{\td\theta}^2\,d\tau^2\Big]
\nonu \\
&+& G^{2/3} \Big[ \fft{3s_{\mu}}{256f}
\left(D\psi+c_{\theta}\,D\phi+ 2fc_{\td\theta} D{\td\phi}\right)^2
+\fft{3s_{\mu}^2\,s_{\theta}^2}{64}\,D\phi^2+
\fft{f}{32g}\,D{\td\phi}^2\Big]\,,
\label{11m111}
\eea
where $G^{-1}=1+{\hat\gamma}^2\,\Delta \ge 1$ and this
geometry is completely regular.

It is also straightforward to apply the above deformation
procedure to the $M^{p,q,r}$ spaces, which have an $SU(3) \times
SU(2) \times U(1)$ isometry and $SO(7)$ holonomy. In particular,
$M^{0,1,0}=\CP^2 \times {\bf S}^3$ has an $SU(3) \times SU(2)^2$
isometry and $M(1,0)={\bf S}^5 \times {\bf S}^2$ has an $SU(4)
\times SU(2)$ isometry \cite{DNP}. However, only the $M^{1,1,1}$
case is supersymmetric.

\subsection{Deforming AdS$_4\times L^{p,q,r,s}$}

We will now consider a class of cohomogeneity three
Sasaki-Einstein spaces $L^{p,q,r,s}$, which include $M^{1,1,1}$
and ${\bf S}^7$ as special cases. The metric for $L^{p,q,r,s}$ is
given by \cite{Cvetic:2005ft}
\bea
ds_{Y_7}^2
& = & \left( d \tau + \sigma \right)^2 +\frac{Y(x)}{4 x F(x)} d
x^2 -\frac{x\left(1-F(x) \right)}{Y(x)} \left( \sum_{i=1}^{3}
\frac{\mu_i^2}{\al_i} d {\varphi}_i  \right)^2
\nonu \\
& + & \sum_{i=1}^{3} \left(1 -\frac{x}{\al_i} \right)
\left(d \mu_i^2 +\mu_i^2 d \varphi_i^2 \right)
 + \frac{x}{\left(\sum_{i=1}^{3} \frac{\mu_i^2}{\al_i}\right)}
\left( \sum_{j=1}^{3} \frac{\mu_j}{\al_j} d \mu_j \right)^2
-\sigma^2,
\label{CLPPmetric}
\eea
where $\sigma, Y(x)$, and $F(x)$ are given by
\bea
\sigma &= & \sum_{i=1}^{3} \left(1-\frac{x}{\al_i} \right)
\mu_i^2 d \varphi_i, \nonu \\
Y(x) & = & \sum_{i=1}^{3} \frac{\mu_i^2}{ \left(\al_i -x\right)},
\qquad F(x) = 1-\left(\frac{\mu}{x}\right) \prod_{i=1}^{3}
\frac{1}{\left(\al_i-x\right)},  \qquad \sum_{i=1}^{3} \mu_i^2
=1. \nonu \eea
This metric smoothly extends onto a complete and non-singular
manifold if $p l_1 + q l_2 + \sum_{j=1}^{3} r_j \frac{\pa}{\pa
\varphi_j}=0$ for coprime integers $(p,q,r_j)$. Note that $p+q
=\sum_{j=1}^{3} r_j$.

We will now discuss the case in which the cohomogeneity one
$X^{p,q}$ spaces of \cite{GMSW03} are recovered. For equal
$\al_i\equiv \alpha$, we can pull the $1/\al_i$ factors of
$\left(1-\frac{x}{\al_i}\right)$ out of the summation symbol in
(\ref{CLPPmetric}). Next, one can write the 5-sphere metric as a
Hopf fibration over $\CP^2$ \cite{GLPP}
\bea \sum_{i=1}^{3} \left( d \mu_i^2 + \mu_i^2 d \varphi_i^2
\right) =\left(d \psi + A \right)^2 + ds_{FS(2)}^2\,, \nonu \eea
where $ ds_{FS(2)}^2$ is the Fubini-Study metric on $\CP^2$ and
$A$ is a local potential for the K\"{a}hler form on $\CP^2$.
Therefore, the three $U(1)$'s are enhanced to $SU(3) \times U(1)$.
Moreover,
\bea \sum_{i=1}^{3} \mu_i^2 d \varphi_i= \left(d \psi +A \right).
\nonu \eea

Upon the coordinate transformation
\bea x \rightarrow y +\frac{1}{3c}, \qquad \mu \rightarrow
\frac{1-a c^2}{3 c^4}, \qquad \al \rightarrow \frac{4}{3c}\,,
\nonu \eea
the second term in (\ref{CLPPmetric}) becomes
\bea \frac{(1-c y)^2}{(\frac{4a}{3}-8 y^2+ \frac{32}{3} c y^2 -4
c^2 y^4) } d y^2 \equiv \frac{d y^2}{\widetilde{F}(y)}\,. \nonu
\eea
Here, $\widetilde{F}(y)$
is the same as the one in (\ref{F})
when $\ell_1=\ell_2$ and $n_1=n_2=c$.
Then the metric can be written as
\be ds_{Y_7}^2 = [d \tau + \ft34 \left(1 -c y \right) \left(d \psi
+A \right)]^2 + \frac{dy^2}{\widetilde{F}(y)} + \frac{9}{64} c^2
{\widetilde{F}}(y) \left(d \psi +A \right)^2 + \frac{3}{4} \left(
1 -c y\right)  ds_{FS(2)}^2\,, \label{Metric} \ee
which is a rescaled version of the cohomogeneity one
generalization of $M^{1,1,1}$ given in \cite{GMSW03}. The original
$U(1)^4$ isometry of 7-dimensional Sasaki-Einstein space is
enhanced to $SU(3) \times U(1)^2$, where the $SU(3)$ symmetry
comes from $\CP^2$ and two $U(1)$'s are parameterized by above
$\tau$ and $\psi$.
The $c=0$ limit reproduces $M^{1,1,1}$ space \cite{GMSW03}.
As observed in \cite{GMSW03}, the limit of
$a=1$ provides the space $\CP^3$ \cite{PP} corresponding to the
last three terms of (\ref{Metric}) with the replacement
$U=c_{\theta}^2$ where the function $U$ is the same as the one in
\cite{GMSW03} and the resulting 7-dimensional space is ${\bf S}^7$
(The structure of the first term of (\ref{Metric}) gives the exact
expression for $U(1)$ bundle).

We will now turn to the deformation of the 11-dimensional solution
given by
\be ds_{11}^2 = \frac{1}{4} ds_{{\rm
AdS}_4}^2+ds_{L^{p,q,r,s}}^2\,, \qquad F_{(4)} = \frac{3}{8}
\left( \frac{\mbox{vol} \; L^{p,q,r,s} } {\mbox{vol}\; {\bf S}^7}
\right)^{1/6} \omega_{AdS_4}\,, \nonu \ee
where the volume of $L^{p,q,r,s}$ is given in
\cite{Cvetic:2005ft}. The three $U(1)$ symmetries which will be
used for the deformation act by shifting the $\varphi_1,
\varphi_2$, and $\varphi_3$ directions. The remaining $U(1)$
symmetry corresponds to the R-symmetry of the dual gauge theory.
As in all of the previous cases, the marginal deformation
preserves the ${\cal N}=2$ supersymmetry.

Since the three-torus ${\bf T}^3$ has coordinates $\varphi_i$
where $i=1,2,3$, it is straightforward to rewrite the above metric
in terms of three-dimensional part and the eight-dimensional part.
As we have done before, by looking at the $d\varphi_3^2$ from
(\ref{CLPPmetric}), we obtain
\bea \Delta^{1/3} e^{4\phi/3} = \frac{\mu_3^2 \left(Y(x) \al_3^2 -
x\left[Y(x) \al_3 -(-1+F(x)) \mu_3^2\right]\right)}{Y(x)
\al_3^2}\,.\label{Deltaphi} \eea
Now we have $\varphi_3$ dependent term as in (\ref{three}) and the
coefficient function $N_1$ and $N_2$ can be determined. The
explicit form for them is given in the Appendix B. By rearranging
the metric, the two-dimensional metric can be written as and the
coefficients $K,L$, and $M$ are presented in the Appendix B.
\bea
 \Delta^{1/3} e^{-2\phi/3} h_{mn} D \varphi^m D \varphi^n
  = K
\left(D \varphi_1 + L D \varphi_2 \right)^2 + \left( \frac{M}{N}
\right) D \varphi_2^2\,. \label{twom111} \eea
from which we find that
\be \Delta=\Big( \fft{KM}{N}\Big)^{3/2}\,e^{2\phi}\,,
\nonu
\ee
which, together with (\ref{Deltaphi}), gives $\Delta$. Finally,
the eight-dimensional metric is written as
\bea
\Delta^{-1/6} g_{\mu \nu} dx^{\mu} dx^{\nu} & =&
\frac{1}{4} ds_{{\rm AdS}_4}^2 +
\frac{Y(x)}{4 x F(x)} d x^2+
 \sum_{i=1}^{3} \left(1 -\frac{x}{\al_i} \right) d \mu_i^2
\nonu \\
&+& \frac{x}{\left(\sum_{i=1}^{3} \frac{\mu_i^2}{\al_i}\right)}
\left( \sum_{j=1}^{3} \frac{\mu_j}{\al_j} d \mu_j \right)^2
+ P  d \tau^2
\label{8metric}
\eea
by collecting the remaining $d \tau^2$ term and the explicit form
for $P$ is in the Appendix B.
After applying an $SL(2,R)$ transformation, we lift back up to
eleven dimensions and find the deformed solution
\bea
ds_{11}^2 = G^{2/3}
\Delta^{1/3} e^{-2\phi/3} h_{mn} D \varphi^m D \varphi^n
+G^{-1/3} \Delta^{-1/6} g_{\mu \nu} dx^{\mu} dx^{\nu},
\label{deformedL} \eea
where
\be G^{-1}=1+{\hat\gamma}^2\Delta\,.
\nonu
\ee
This is a completely regular geometry.

Note that the deformation of AdS$_4\times M^{1,1,1}$ discussed in
the previous subsection cannot be obtained as a special case of
(\ref{deformedL}). This is because the $M^{1,1,1}$ metric
(\ref{11m111})
was expressed in different coordinates, which
means that we performed the dimensional reductions along different
directions. However, the deformation of AdS$_4\times {\bf S}^7$
arises from (\ref{deformedL}) in the limit of equal $\alpha_i$ and
$a=1$.

\section{Further directions}

We have generated deformations based on $U(1)^3$ symmetries of
11-dimensional geometries which involve various 7-dimensional
Sasaki-Einstein spaces. The initial geometries have the form of a
direct product of AdS$_4$ and a 7-dimensional Sasaki-Einstein
space $Y_7$, and the deformed geometries are a warped product of
these spaces. The warp factor depends on three of the internal
directions of $Y_7$. In fact, supersymmetric AdS in warped
spacetimes were previously obtained from direct products of AdS
and an internal space \cite{warpedAdS}. Furthermore, the
corresponding consistent Kaluza-Klein warped embeddings of gauged
supergravities were found \cite{warpedKK}. That is, the vacuum AdS
solutions of these gauged supergravities give rise to warped
products with internal spaces. However, in most of these cases,
the warp factor (depends on only one of the internal directions)
is singular. There are some cases which have a non-singular warp
factor but then the internal space has orbifold-type conical
singularities. The warped spacetimes discussed in \cite{LM} and in
this paper are completely regular. Therefore, it would be
interesting to see if one could construct consistent Kaluza-Klein
warped embeddings of gauged supergravities whose AdS vacua give
rise to some of these solutions. In particular, one might be able
to deform previously-known sphere reductions whose initial vacuum
geometries are of the form AdS$_n\times {\bf S}^m$.

The general class of Sasaki-Einstein spaces discussed in section 3
include ${\bf S}^7$ and $Q^{1,1,1}$ as special cases. The most
interesting problem is what are the dual conformal field theories
corresponding to M-theory on $AdS_4 \times Y_7$. For
$Y_7=Q^{1,1,1}$ space, it is known that the gauge theory is
$SU(N)^3$ with three chiral superfields \cite{Fabbri2}. The field
contents are represented by a quiver diagram where nodes represent
the gauge groups and link matter fields in the bi-fundamental
representation of the gauge group they connect. It would be
interesting to construct the complete spectrum of 11-dimensional
supergravity compactified on $Y_7$ and compare the Kaluza-Klein
spectrum with that of the corresponding gauge theory in the large
$N$ limit.

Most probably, the classes of 7-dimensional Sasaki-Einstein spaces
discussed in this paper do not exhaust all of the possibilities.
For instance, we have noted that ${\bf S}^7$ arises as a special
case of two separate branches of cohomogeneity one families of
Sasaki-Einstein spaces $X^{p,q}$, one with a 4-dimensional base
space $B_4=\CP^2$ and the other with $B_4=\CP^1\times \CP^1$. The
question arises as to whether these two branches can be
encompassed by a more general family of Sasaki-Einstein spaces.

On a related note, the generalization of such spaces to larger
cohomogeneity tends to involve breaking $SU(2)$ and $SU(3)$
elements of the isometry group into $U(1)$ and $U(1)^2$ subgroups,
respectively. For example, the isometry group of $M^{1,1,1}$ is
$SU(3)\times SU(2)\times U(1)$. The cohomogeneity one
generalization $X^{p,q}$ with $B_4=\CP^2$ involves replacing the
$SU(2)$ with $U(1)$, and the further generalization to
cohomogeneity three $L^{p,q,r,s}$ involves replacing the $SU(3)$
with $U(1)^2$. Thus, the fact that the isometry group of
$Z^{p,q,r,s}$ contains two $SU(2)$ factors implies that there may
be room for further generalization.

One can also apply the deformation procedure to cases for which do
not involve a Sasaki-Einstein space. For example, one might
consider geometries of the form AdS$_4\times M^7$, where $M^7$ is
a 3-Sasakian or proper weak $G_2$ manifold. These are gravity
duals of ${\cal N}=3$ and ${\cal N}=1$ superconformal field
theories, respectively. An infinite family of $M^7$ spaces have
been constructed in \cite{sakaguchi} as principle $SO(3)$ bundles
over Tod-Hitchin metrics. These spaces are parameterized by a
single integer. A particular case of this family of spaces is
$N^{0,1,0}$ \cite{CR,PP84,Castellani,AR99,Ahn02}.
The $SU(3)\times SU(2)$ isometry of $N^{0,1,0}$
contains $U(1)^3$ as a subgroup, which can be used in the
deformation procedure. The dual gauge theory has ${\cal N}=3$ or
${\cal N}=1$ supersymmetry, depending on the orientation of the
space. We expect some of the supersymmetry to be broken by the
deformation in the ${\cal N}=3$ case but not in the ${\cal N}=1$
case.

Another special case which is a proper weak $G_2$ manifold
is the squashed 7-sphere \cite{ADP,DNP1,DNP}, which we denote as
${\widetilde {\bf S}}^7$. This space can be described as a
squashed $SO(3)$ bundle over a 4-sphere which is entirely
equivalent to the standard Fubini-Study metric \cite{DNP}. In this
scheme the squashing corresponds to changing the size of the
$SO(3)$ fiber relative to the ${\bf S}^4$ base space. The
AdS$_4\times {\widetilde {\bf S}}^7$ solution of 11-dimensional
supergravity is expected to be dual to a 3-dimensional ${\cal
N}=1$ gauge theory with $SO(5)\times SO(3)$ global symmetry
\cite{AR} for the squashing with left-handed orientation. One can
apply the techniques for generating marginal deformations by
choosing two $U(1)$'s from $SO(5)$ and one from $SO(3)$.

One could also apply these deformations to gravity duals of field
theories which exhibit RG flows away from a conformal fixed-point.
This has already been done in \cite{LM} for a fractional D3-brane
on a deformed conifold \cite{strassler}, which corresponds to a
4-dimensional ${\cal N}=1$ gauge theory which flows from a UV
conformal fixed-point to a confining theory in the IR region.
Also, \cite{GN} has studied the deformations of a D5-brane wrapped
on an ${\bf S}^2$ of a resolved conifold, which is another example
of a gravity dual of an ${\cal N}=1$ gauge theory \cite{nunez}.
There are also regular 11-dimensional solutions which exhibit
flows from the geometries we have discussed. For example, the
near-horizon region of certain deformed M2-branes have geometries
which go from AdS$_4\times Q^{1,1,1}$ or AdS$_4\times M^{1,1,1}$
to other smooth geometries \cite{gibbons}. The 3-dimensional gauge
theory interpretation of these solutions is that there is an RG
flow from a UV conformal fixed-point to a confining gauge theory.
Marginal deformations can be applied to these theories as well.

\vspace{1cm}
\centerline{\bf Acknowledgments}

\indent

We would like to thank Hong L\"{u}, Oleg Lunin, Juan Maldacena,
and James Sparks for useful discussions. In particular, we
appreciate the detailed explanations provided by Oleg Lunin with
regards to his recent paper. The work of C.A. is supported by a
grant from the Monell Foundation through the Institute for
Advanced Study,
and by Korea Research Foundation Grant KRF-2002-015-CS0006. The
work of J.F.V.P. is supported by DOE grant DOE-FG02-84ER-40153.

\appendix

\renewcommand{\thesection}{\large \bf \mbox{Appendix~}\Alph{section}}
\renewcommand{\theequation}{\Alph{section}\mbox{.}\arabic{equation}}

\section{Compendium of Sasaki-Einstein spaces}

\begin{figure}
   \epsfxsize=4.0in
   \centerline{\epsffile{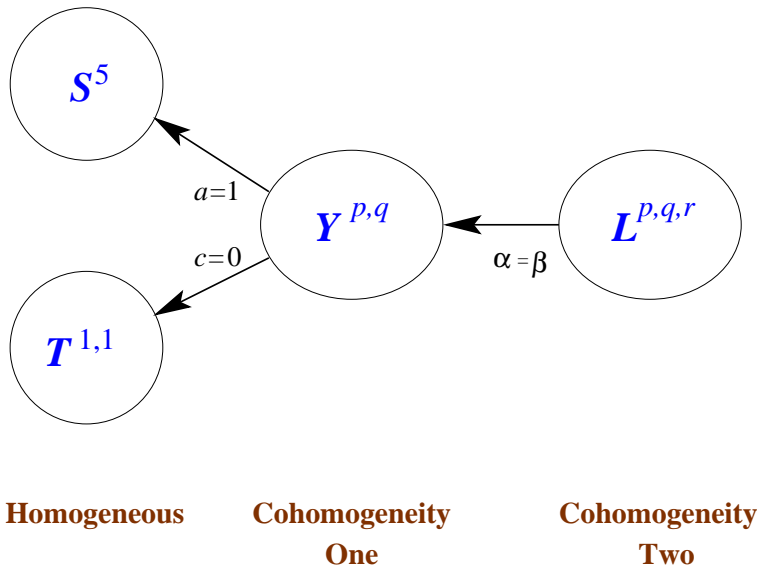}}
   \caption[FIG. \arabic{figure}.]{Flow diagram for 5-dimensional Sasaki-Einstein spaces}
\end{figure}

Until recently, the only explicitly known 5-dimensional
Sasaki-Einstein metrics were the round metric on ${\bf S}^5$, the
homogeneous metric on $T^{1,1}$ and their quotients. We now know
of a countably infinite number of explicit cohomogeneity one
5-dimensional Sasaki-Einstein metrics $Y^{p,q}$
\cite{GMSW03-1,GMSW03-2}. These spaces are specified by a single
nontrivial parameter, though it is useful to keep the two
parameters $a$ and $c$ in order to clearly see the special cases.
In particular, $c=0$ corresponds to the limit in which the space
is $T^{1,1}$. In this case, $a$ is a trivial rescaling parameter.
On the other hand, the space is ${\bf S}^5$
for $a=1$, in which case $c$
is now a trivial rescaling parameter. The $Y^{p,q}$ spaces can be
characterized by two relatively prime positive integers $p$ and
$q$. This class of spaces has been further generalized by taking
the BPS limits of Euclideanized Kerr-de Sitter black hole metrics
with two independent angular momenta parameters $\alpha$ and
$\beta$ \cite{Cvetic:2005ft}. The resulting cohomogeneity two
metrics $L^{p,q,r}$ are characterized by the positive coprime
integers $p$, $q$ and $r$. The $Y^{p,q}$ spaces lie in the
subclass of the $L^{p,q,r}$ spaces for which $\alpha=\beta$. The
flow diagram in Figure 1 shows how all of the above 5-dimensional
Sasaki-Einstein spaces are related. The $L^{p,q,r}$ metric is
explicitly written in Appendix C.

The cohomogeneity two $L^{p,q,r}$ spaces generally have $U(1)^3$
isometry. For the subclass of spaces with cohomogeneity less than
two, a $U(1)$ or $U(1)^2$ element of the isometry group gets
enhanced to $SU(2)$ or $SU(3)$, respectively. In particular, when
$\alpha=\beta$, we get the cohomogeneity one $Y^{p,q}$ spaces with
$SU(2)\times U(1)^2$ isometry. The further limit $c=0$ yields the
homogeneous space $T^{1,1}$, which has $SU(2)^2\times U(1)$
isometry. Note that the remaining $U(1)$ element corresponds to
the R-symmetry of the dual field theory. The exception to the
above is ${\bf S}^5$, whose isometry group is $SO(6)$. In order to
perform the type of deformation discussed in \cite{LM}, it is
crucial that the isometry group of all of these spaces contains
$U(1)^3$.

The higher-dimensional analog of the $Y^{p,q}$ spaces was found in
\cite{GMSW03}. Like the $Y^{p,q}$ spaces, these are also specified
by a single nontrivial parameter which can be written in terms of
two relatively prime positive integers $p$ and $q$. Again, we will
keep two parameters $a$ and $c$ in order to study special cases.
We focus on the 7-dimensional spaces, which we refer to as
$X^{p,q}$. Note that the 5-dimensional $Y^{p,q}$ spaces could be
expressed as a $U(1)$ bundle over a 4-dimensional
Einstein-K\"{a}hler space which, in turn, was a 2-bundle over
${\bf S}^2$. Similarly, the 7-dimensional $X^{p,q}$ can be
expressed as a $U(1)$ bundle over a 6-dimensional
Einstein-K\"{a}hler space which is itself a 2-bundle over a
4-dimensional Einstein-K\"{a}hler space $B_4$. We have a couple of
choices for the base space $B_4$, namely $\CP^2$ or $\CP^1\times
\CP^1$.

We will first consider $B_4=\CP^2$, which is quite analogous to
the 5-dimensional case. $c=0$ corresponds to the space
$M^{1,1,1}$, for which $a$ becomes the trivial rescaling
parameter. $a=1$ corresponds to the space ${\bf S}^7$, and now $c$
is the trivial rescaling parameter. These $X^{p,q}$ spaces can be
generalized to cohomogeneity three spaces $L^{p,q,r,s}$
\cite{Cvetic:2005ft}, which can be found from the BPS limit of
7-dimensional Euclideanized Kerr-de Sitter black holes which have
three independent angular momenta $\alpha_i$. The $X^{p,q}$
subclass of these spaces corresponds to all three angular momenta
$\alpha_i$ being equal. The right-hand portion of Figure 2
describes how these spaces are related. There is also an
intermediate subclass which we have not included explicitly in
Figure 2. Namely, there is a subclass of the $L^{p,q,r,s}$ spaces
with cohomogeneity two, for which only two of the $\alpha_i$ are
equal\footnote{We thank Hong L\"{u} for clarifying this point.}.

We now turn to the case of $B_4=\CP^1\times \CP^1$. This family of
$X^{p,q}$ spaces also has a special $c=0$ limit, in which case we
have $Q^{1,1,1}$. However, $a=1$ does not correspond to ${\bf
S}^7$. In fact, we have not been able to identify this case with
any previously-known Sasaki-Einstein space, though it does share a
similarity with ${\bf S}^7$. In particular, like the Fubini-Study
metric on $\CP^3$, the Einstein-K\"{a}hler base space is foliated
by $T^{1,1}$. However, each ${\bf S}^2$ has an identical metric
factor. In fact, this brand of $X^{p,q}$ can be further
generalized by rendering the characteristic radii of the spheres
to be different \cite{GMSW11,CLPP}. This yields a more general
cohomogeneity one family of Sasaki-Einstein spaces $Z^{p,q,r,s}$,
which are characterized by four integers $p,q,r,s$. These spaces
can be written in terms of four parameters $\ell_1$, $\ell_2$,
$n_1$ and $n_2$, although only $\ell_1$ and $\ell_2$ are
nontrivial parameters. This is because the $n_i$ can be rescaled
(\ref{rescaling})
such that only the ratio $n_1/n_2$ is significant. Also, in order
for the $Z^{p,q,r,s}$ spaces to be Einstein, this ratio is written
in terms of the $\ell_i$ (\ref{constraint}).
Nevertheless, it is useful to express
the metric in terms of $n_i$, in order to see the special cases
more clearly. In particular, when $\ell_i=1$ and $n_1=n_2$,
we recover the $X^{p,q}$ spaces. On the other hand, when the
$\ell_i=1$ but $n_1=-n_2$, we actually recover ${\bf
S}^7$ by (\ref{CP3}) and (\ref{Kahler}).
Therefore, the $Z^{p,q,r,s}$ spaces are needed in order to
encompass $Q^{1,1,1}$ (\ref{Qlimit})
and ${\bf S}^7$ within a single family of
cohomogeneity one Sasaki-Einstein spaces. The left-hand side of
Figure 2 describe the various special cases of the $Z^{p,q,r,s}$
family of spaces.

\begin{figure}
   \epsfxsize=6.0in
   \centerline{\epsffile{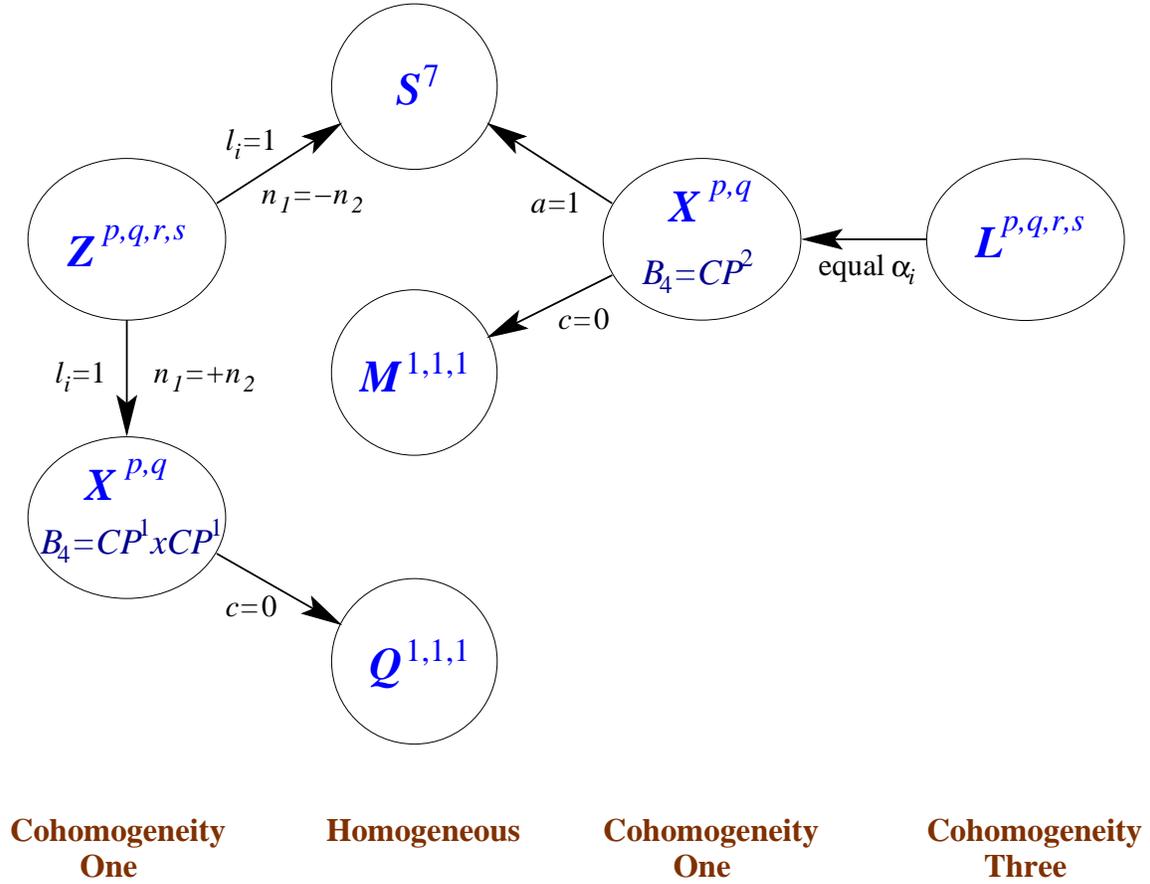}}
   \vspace{.3cm}
   \caption[FIG. \arabic{figure}.]{Flow diagram for 7-dimensional Sasaki-Einstein spaces}
\end{figure}

In general, the cohomogeneity three $L^{p,q,r,s}$ spaces have
$U(1)^4$ isometry. This is enhanced to $SU(3)\times U(1)^2$
isometry for the $X^{p,q}$ spaces with $B_4=\CP^2$, and is further
enhanced to $SU(3)\times SU(2)\times U(1)$ for $M^{1,1,1}$. Since
both the $Z^{p,q,r,s}$ spaces and $X^{p,q}$ with $B_4=\CP^1\times
\CP^1$ have cohomogeneity one, it is not surprising that they
share $SU(2)^2\times U(1)^2$ isometry. This is enhanced to
$SU(2)^3\times U(1)$ for the case of $Q^{1,1,1}$. The exception to
the above is ${\bf S}^7$, which has $SO(8)$ isometry. All of the
above 7-dimensional spaces have isometry groups which contain
$U(1)^4$, which is necessary in order to perform deformations
using the $U(1)^3$ global symmetry while leaving the $U(1)$
R-symmetry untouched.

\section{\large \bf The details for
cohomogeneity three generalization of $M^{1,1,1}$}

For convenience, we list all the coefficient functions
$N_1,N_2,K,L,M,N,P$ and gauge connections ${\cal A}^i$.
The coefficient functions $N_1$ and $N_2$ are
\bea
N_1 & = & -\frac{\left(-1+F\right) x
\al_3 \mu_1^2}{\al_1\left(-Y \al_3^2 +
x\left[Y \al_3 -\left(-1+F\right)
\mu_3^2\right]\right)},
\nonu \\
N_2  & = &   -\frac{\left(-1+F\right)
x \al_3 \mu_2^2}{\al_2\left(-Y \al_3^2 +
x\left[Y \al_3 -\left(-1+F\right)
\mu_3^2\right]\right)}.
\nonu
\eea
The coefficient functions $K,L,M$, and $N$
appearing in (\ref{twom111})
can be obtained as follows:
\bea
K & = &
 \frac{- \Delta^{1/3} e^{4\phi/3}
N_1^2 Y \al_1^2 + \mu_1^2 \left(Y \al_1^2 - x\left[Y \al_1 -\left(
-1+F\right) \mu_1^2\right]\right)}{Y \al_1^2},
\nonu \\
L & = &
\frac{\al_1\left[\Delta^{1/3} e^{4\phi/3}
N_1 N_2 Y \al_1 \al_2 -\left(-1+F\right) x \mu_1^2 \mu_2^2\right] }
{\al_2\left[\Delta^{1/3} e^{4\phi/3} N_1^2 Y \al_1^2 +\mu_1^2
\left(-Y \al_1^2+
x\left[Y \al_1-\left(-1+F\right) \mu_1^2\right]\right)\right]},
\nonu \\
M & = & \Delta^{1/3} e^{4\phi/3}\left[ N_2^2 \al_2^2 \mu_1^2
\left(Y \al_1^2 - x \left[ Y \al_1 -\left(-1+F\right) \mu_1^2\right]
\right)-2 N_1 N_2 \left(-1+F\right) x \al_1
\al_2 \mu_1^2 \mu_2^2 \right.
\nonu \\
&+ &  \left. N_1^2 \al_1^2 \mu_2^2 \left(Y \al_2^2 -x \left[
Y \al_2 -\left(-1+F\right) \mu_2^2\right] \right)\right]
\nonu \\
&+ &  \mu_1^2 \mu_2^2 \left( -Y \al_1^2 \al_2^2 -
x^2 \left[ Y \al_1 \al_2 -\left(-1+F\right)
\left(\al_2 \mu_1^2 +\al_1 \mu_2^2\right) \right]
\right. \nonu \\
&+ & \left.
x\left[Y \al_1 \al_2
\left(\al_1 +\al_2\right) -\left(-1+F\right)\left
(\al_2^2 \mu_1^2 +\al_1^2
\mu_2^2\right)\right]\right),
\nonu \\
N & = & \al_2^2\left[\Delta^{1/3} e^{4\phi/3} N_1^2 Y \al_1^2 +
\mu_1^2\left(-Y \al_1^2 + x \left[ Y\al_1 -\left(-1+F\right) \mu_1^2
\right]\right)\right].
\nonu
\eea
The coefficient function $P$ appearing in (\ref{8metric})
is given by
\bea
P & = & \frac{1}{KM\Delta^{1/3} e^{4\phi/3} \al_1^2 \al_2^2 \al_3^2}
\left(- M \Delta^{1/3} e^{4\phi/3} \al_2^2 \left[
\al_1 \al_3\left(-\mu_1^2
+ N_1 \mu_3^2\right)+x \left(
\al_3 \mu_1^2 - N_1 \al_1 \mu_3^2\right)\right]^2
\right.
\nonu \\
&+ &
K\left(M \al_1^2 \al_2^2 \left[ \Delta^{1/3} e^{4\phi/3}
\al_3^2 -\left(x-
\al_3\right)^2\mu_3^4\right] -
\Delta^{1/3} e^{4\phi/3} \left( L \al_2 \left[\al_1
\al_3 \left(-\mu_1^2 + N_1 \mu_3^2\right)
\right. \right. \right.
\nonu \\
&+ &  \left. \left. \left. \left. x (\al_3 \mu_1^2 - N_1
\al_1 \mu_3^2) \right] +
\al_1 [\al_2 \al_3 (\mu_2^2 -
N_2 \mu_3^2)+
x (-\al_3 \mu_2^2 + N_2 \al_2 \mu_3^2)
]\right)^2 \right) \right).
\nonu
\eea
The gauge connections are given by
\bea
{\cal A}^1 &=&
\frac{1}{KM} \left[-\frac{\left(KL^2+M\right)
\left(x-\al_1\right)\mu_1^2}{\al_1} +
\frac{1}{\al_2 \al_3}\left(M N_1 \al_2
\left(x-\al_3\right)\mu_3^2 \right. \right.
\nonu \\
&+&\left. \left.
KL\left[ -\al_2 \al_3 \left(\mu_2^2 +\left(LN_1
      -N_2\right)\mu_3^2\right)+ x
\left(\al_3\mu_2^2 +
\left(LN_1 -N_2\right)\al_2
\mu_3^2\right)\right] \right) \right],
\nonu \\
{\cal A}^2 &=& \frac{1}{M} \left[ \left(1 -\frac{x}{\al_2} \right)
\mu_2^2 + \frac{N_2\left(x-\al_3\right) \mu_3^2}{\al_3} +L
\left(\left(-1+\frac{x}{\al_1}\right)\mu_1^2+
\frac{N_1\left(-x+\al_3\right)\mu_3^2}{\al_3} \right)\right],
\nonu
\eea
\bea
{\cal A}^3 &=& -\frac{1}{\Delta^{1/3} e^{4\phi/3}} \left(
-\mu_3^2+\frac{x \mu_3^2}{\al_3} + \frac{\Delta^{1/3} e^{4\phi/3}
N_2}{M}\left[ \left(1-\frac{x}{\al_2}\right) \mu_2^2
+\frac{N_2\left(x-\al_3 \right)\mu_3^2}{\al_3}
\right. \right. \nonu \\
&+& \left.
L\left(\left(-1+\frac{x}{\al_1} \right)\mu_1^2+
\frac{N_1\left(-x+\al_3\right)\mu_3^2}{\al_3}
\right)\right] \nonu \\
&+&
\frac{1}{KM} \left[\Delta^{1/3} e^{4\phi/3} N_1\left(
-\frac{(KL^2+M)(x-\al_1)\mu_1^2}{\al_1} \right. \right. \nonu \\
&+ & \frac{1}{\al_2 \al_3}
\left[ MN_1 \al_2\left(x-\al_3\right)\mu_3^2+KL\left(-\al_2\al_3
\left(\mu_2^2 +
\left(LN_1-N_2\right)\mu_3^2\right) \right. \right. \nonu \\
&+& \left. \left. \left. \left. \left. x \left[\al_3\mu_2^2
+\left(LN_1-N_2\right)\al_2 \mu_3^2\right] \right)\right] \right)
\right] \right).
 \nonu \eea

\section{\large \bf Deforming AdS$_5\times L^{p,q,r}$}

We have focused on deformations of supergravity solutions with
$U(1)^3$ global symmetry. Here we note that an analogous procedure
can also be applied to the cohomogeneity two 5-dimensional
Sasaki-Einstein space $L^{p,q,r}$ found in \cite{Cvetic:2005ft}. A
solution of type IIB supergravity is given by
\be ds_{10}^2=ds_{{\rm AdS}_5}^2+ds_{L^{p,q,r}}^2\,,\qquad F_{(5)}
= \fft{16\pi N}{{\cal V}}\,(\omega_{{\rm AdS}_5}+\ast \omega_{{\rm AdS}_5})\,,
\nonu
\ee
where $N$ is the number of D3-branes and ${\cal
V}=\mbox{vol}(L^{p,q,r})/\mbox{vol}({\bf S}^5)$, which is generically irrational
and less than one. The volume of $L^{p,q,r}$ is given in
\cite{Cvetic:2005ft}. The five-dimensional Sasaki-Einstein space
$L^{p,q,r}$ has an isometry $U(1)^3$ and is given by
\bea
ds_{L^{p,q,r}}^2 &=& (d\tau+\sigma)^2
+\fft{\rho^2}{4\Delta_x}\,dx^2
+\fft{\rho^2}{\Delta_{\theta}}\,d\theta^2
+\fft{\Delta_x}{\rho^2}\, \Big( \fft{s_{\theta}^2}{\alpha}d\phi+
\fft{c_{\theta}^2}{\beta}d\psi \Big)^2\nnn\\
&+& \fft{\Delta_{\theta}s_{\theta}^2\,c_{\theta}^2}{\rho^2}\,
\Big[ \left( 1-\fft{x}{\alpha} \right)
d\phi-\left( 1-\fft{x}{\beta} \right) d\psi
\Big]^2\,,
\nonu
\eea
where
\bea
\sigma &=& \left( 1-\fft{x}{\alpha} \right)
s_{\theta}^2\,d\phi+
\left( 1-\fft{x}{\beta} \right)
c_{\theta}^2\,d\psi\,,\nnn\\
\Delta_x &=& x(\alpha-x)(\beta-x)-1\,,\qquad
\rho^2=\Delta_{\theta}-x\,,\nnn\\ \Delta_{\theta} &=&
\alpha\,c_{\theta}^2+\beta\,s_{\theta}^2\,.
\nonu
\eea
We put $\mu=1$ in the last term of $\Delta_x$.
When $p=q=r=1$(according to \cite{Cvetic:2005ft}, $\alpha=\beta$),
implying $SU(2)^2 \times U(1)$ symmetry,
the metric becomes $T^{1,1}$ space and When
$\mu=0$, one obtains round 5-sphere.

The two $U(1)\times U(1)$ symmetry is associated with shifts in
$\phi$ and $\psi$. In order to make the ${\bf T}^2$
more explicit, we
express the metric as
\bea
ds_{10}^2 &=& ds_{{\rm
AdS}_5}^2+\fft{\rho^2}{4\Delta_x}\,dx^2
+\fft{\rho^2}{\Delta_{\theta}}\,d\theta^2+N\,d\tau^2
+\fft{c_{\theta}^2}{\beta^2 \rho^2 L}\,
(d\psi+\beta\rho^2LM\,d\tau )^2 \nnn\\ &+&
\fft{s_{\theta}^2}{\alpha^2\rho^2K}\, \Big(
d\phi-\fft{\alpha\rho}{\beta} c_{\theta}^2\,K\,d\psi+
\alpha \rho^2
(\alpha-x)K\,d\tau\Big) ^2 \,,
\nonu
\eea
where the coefficient functions are
\bea
K^{-1} &=& (\alpha-x)^2(\Delta_{\theta}-x\,s_{\theta}^2)
+\Delta_x s_{\theta}^2\,,\nnn\\
L^{-1} &=& (\beta-x)^2 (\Delta_{\theta}-x\,c_{\theta}^2) +\Delta_x
c_{\theta}^2-\rho^2 K\, s_{\theta}^2\,c_{\theta}^2\,,\nnn\\
M &=& (\beta-x)+(\alpha-x)\rho K\,s_{\theta}^2\,,\nnn\\
N &=& 1-(\alpha-x)^2\rho^2 K\,s_{\theta}^2- \rho^2
LM^2\,c_{\theta}^2\,.
\nonu
\eea
The deformed metric is given by
\bea ds_{10}^2 &=& G^{-1/4} \Big[ ds_{{\rm
AdS}_5}^2+\fft{\rho^2}{4\Delta_x}\,dx^2
+\fft{\rho^2}{\Delta_{\theta}}\,d\theta^2+N\,d\tau^2\Big] +
G^{3/4}\Big[ \fft{c_{\theta}^2}{\beta^2 \rho^2 L}\,
(d\psi+\beta\rho^2LM\,d\tau )^2 \nnn\\ &+&
\fft{s_{\theta}^2}{\alpha^2\rho^2K}\, \Big(
d\phi-\fft{\alpha\rho}{\beta} c_{\theta}^2\,K\,d\psi+\alpha \rho^2
(\alpha-x)K\,d\tau\Big) ^2\Big] \,, \nonu \eea
where
\be G^{-1}=1+{\hat\gamma}^2 \fft{s_{\theta}^2
c_{\theta}^2}{\alpha^2 \beta^2 \rho^4 LK}\,.
\nonu
\ee

\end{document}